\documentclass[fleqn,usenatbib]{mnras}

\def\bri{{\beta_{\rm reio}}}
\def\bcmb{{\beta_{\rm recomb}}}
\def\bhi{{\theta_{\rm recomb}}}
\def\blo{{\theta_{\rm reio}}}
\def\bdelt{{\Delta \beta}}
\usepackage{amsmath,amssymb,bm,color,Macro}
\def\tn#1{\textcolor{black}{#1}}
\def\bds#1{\textcolor{black}{#1}}

\usepackage[T1]{fontenc}

\DeclareRobustCommand{\VAN}[3]{#2}
\let\VANthebibliography\thebibliography
\def\thebibliography{\DeclareRobustCommand{\VAN}[3]{##3}\VANthebibliography}

\usepackage{graphicx}	
\usepackage{amsmath}	
\usepackage{amssymb}	

\usepackage{newtxtext,newtxmath}

\title[Cosmic birefringence from reionization signals]{Cosmic birefringence tomography and calibration-independence with reionization signals in the CMB}

\author[Blake D. Sherwin \& Toshiya Namikawa]{
Blake D. Sherwin$^{1,2}$\thanks{E-mail: sherwin@damtp.cam.ac.uk} and
Toshiya Namikawa$^{3,1}$\thanks{E-mail: tn334@cam.ac.uk}
\\
$^{1}$Department of Applied Mathematics and Theoretical Physics, University of Cambridge, Wilberforce Road, Cambridge CB3 0WA, United Kingdom \\
$^{2}$Kavli Institute for Cosmology, University of Cambridge, Madingley Road, Cambridge CB3 OHA, United Kingdom \\
$^3$Kavli Institute for the Physics and Mathematics of the Universe (WPI), UTIAS, The University of Tokyo, Kashiwa, Chiba 277-8583, Japan
}

\date{Accepted XXX. Received YYY; in original form ZZZ}

\pubyear{2021}

\begin{document}
\label{firstpage}
\pagerange{\pageref{firstpage}--\pageref{lastpage}}
\maketitle

\begin{abstract}
The search for cosmic polarization rotation or birefringence in the CMB is well-motivated because it can provide powerful constraints on parity-violating new physics, such as axion-like particles. In this paper we point out that since the CMB polarization is produced at two very different redshifts -- it is generated at both reionization and recombination -- new parity-violating physics can generically rotate the polarization signals from these different sources by different amounts. We explore two implications of this. First, measurements of CMB birefringence are challenging because the effect is degenerate with a miscalibration of CMB polarization angles; however, by taking the difference of the reionization and recombination birefringence angles (measured from different CMB angular scales), we can obtain a cosmological signal that is immune to instrumental angle miscalibration. Second, we note that the combination with other methods for probing birefringence can give tomographic information, constraining the redshift origin of any physics producing birefringence. We forecast that the difference of the reionization and recombination birefringence angles can be competitively determined to within $\sim 0.05$ degrees for future CMB satellites such as LiteBIRD. Although much further work is needed, we argue that foreground mitigation for this measurement should be less challenging than for inflationary B-mode searches on similar scales due to larger signals and lower foregrounds. 
\end{abstract}




\section{Introduction and approach}

Precise measurements of the cosmic microwave background (CMB) anisotropies have provided us with a wealth of information on the current standard cosmological model. However, CMB polarization signals -- particularly the $B$-modes, an odd-parity twisting pattern in the polarization map -- are still dominated by instrumental noise over a wide range of angular scales. Upcoming CMB experiments such as the BICEP Array \citep{BICEPArray}, Simons Observatory \citep{SimonsObservatory}, CMB-S4 \citep{CMBS4} and LiteBIRD \citep{LiteBIRD}, with which the polarization noise will be reduced significantly, are therefore expected to make significant scientific advances. 

One effect that can be probed with high-precision $B$-modes is {\it cosmic birefringence}: a rotation of the linear polarization angle of the CMB during the propagation from last scattering to the observer. \bds{This rotation by an angle $\beta$ converts $E-$modes to $B-$modes, giving an observed $B-$mode (absent primordial $B$) of $B'_{lm} = E_{lm}\sin2 \beta$. As this produces a new, non-zero correlation between the observed $E$- and $B$-modes, high precision measurements of the $EB$ cross-power spectrum can provide tight constraints on cosmic birefringence} \footnote{Cosmic birefringence also introduces a temperature-$B$ correlation, but this spectrum is a less sensitive probe of birefringence than the $EB$ spectrum \citep{Keating:2013,P16:rot}. Therefore, in this paper, we only consider the $EB$ spectrum. 
}. 

Several types of beyond-the-Standard-Model physics can source cosmic birefringence, with birefringence typically caused by parity-violating interactions. In this paper, although our method is generally applicable, we focus as an illustrative example on the cosmic birefringence induced by axion-like particles (ALPs) that couple to photons through a so-called Chern-Simons term in the Lagrangian (see e.g., \citealt{Carroll:1998,Liu:2006:biref,Li:2008,Pospelov:2009,Finelli:2009,Liu:2016:axion-DM,Hlozek:2017:axion,Fujita:2020:biref,Fujita:2020:isobiref,Takahashi:2020:biref} and a review, \citealt{Marsh:2016}): 
\begin{equation}
    \mS{L}\supset \frac{g_{a\gamma}}{4} a F_{\mu\nu}\tilde{F}^{\mu\nu} 
    \,,
\end{equation}
where $a$ is the ALP field, $F_{\mu\nu}$ is the electromagnetic field, $\tilde{F}^{\mu\nu}$ is its dual, and $g_{\alpha\gamma}$ is the coupling constant between ALPs and electromagnetic fields. 
The existence of such ALPs is a generic prediction of string theory; \tn{
in addition,  
birefringence-inducing ALPs could be candidates for an early dark energy mechanism that aims to resolve the current Hubble parameter tension \citep{Capparelli:2019:CB}. Furthermore, a multi-field model for ALPs simultaneously predicts cosmic birefringence and the existence of the dark matter \citep{Obata:2021}. 
}

The ALPs introduce a polarization angle rotation of $\beta = g_{a\gamma}\Delta a/2$, where $\Delta a$ is the change in the field value $a$ over the photon trajectory \citep{Carroll:1989:rot,Harari:1992:axion}. 
If we only consider a time-dependent background evolution of the field $a$ (neglecting any spatial variation), the result is isotropic birefringence: the rotation of polarization by the same angle, irrespective of the observation direction in the sky. (We will only consider isotropic birefringence in this paper; anisotropic birefringence induced by ALPs is discussed in several other publications such as \citealt{Kamionkowski:2010:AnisoCB,Namikawa:2020:biref,Bianchini:2021:biref}). 

\bds{The ALP field is initially nearly constant, but it begins to oscillate when $H(z)\sim m$; the ALPs' energy density then dilutes and their field value falls quickly (see, e.g., \citealt{Marsh:2016}). Assuming that the ALPs begin to oscillate at some time between CMB recombination and the observation time today, we expect non-zero $\beta = \bcmb$ due to the change of the ALPs' field value over the photons path.}

In this paper we note that since the reionization CMB polarization signal arises from a much lower redshift $z\sim 8$ than the primary CMB polarization ($z \sim 1100$), the birefingence angle for reionization polarization $\bri$ may in general be different from the birefringence angle for recombination polarization $\bcmb$. Indeed, if we consider ALPs of mass $m$ as a source of birefringence, $\bri$ is similar to $\bcmb$ only in the case that $H$ becomes $ \sim m$ well after $z \sim 8$; otherwise we expect that $\bri \ll \bcmb$ if $\bcmb$ is non-zero. \bds{(We note that in LCDM $H$ increases by only one order of magnitude from $z \approx 0-8$, but by three orders of magnitude from $z \approx 8-1100$; therefore, there is a large range of masses for which $ H \sim m$ before reionization such that $\bri \ll \bcmb$; of course, beyond such simple arguments, predictions for birefringence are model-dependent.}) 

The fact that we may quite generically expect that $\bri \neq \bcmb$ -- or equivalently that the birefringence difference  $\bdelt \equiv \bcmb-\bri$ is non-zero for many typical models that produce cosmic birefringence -- has two important cosmological applications.

\emph{1. Angle-calibration degeneracy-breaking:} Probing $\bcmb$
via the $EB$ power spectrum is complicated by the fact that the effect is degenerate with CMB experiment angle miscalibration, which can similarly rotate polarization by an amount $\alpha$ \citep{QUaD:2008:biref,Miller:2009:biref,WMAP7,Keating:2013} \footnote{Several other instrumental systematics could also produce biases in $EB$ measurements, although such effects can be cross-checked by an alternative way of measuring the polarization angle recently proposed by \citet{Namikawa:2021:MC} for low-noise experiments. Furthermore, \citet{Abitbol:2016:eb} pointed out that Galactic foregrounds could bias estimates of the polarization angle.}; the $EB$ power spectrum at intermediate and high multipoles $l$ therefore naively only constrains the combination $\bhi = \alpha+\bcmb$. Several experiments have placed constraints on $\bhi$ using the $EB$ power spectrum. Some of the results show a detection of non-zero $EB$ correlation (see, e.g., \citealt{P16:rot}) while a recent ACT analysis finds that the $EB$ spectrum is consistent with zero within $2\sigma$ \citep{ACT:Choi:2020,Namikawa:2020:biref}. However, we cannot reliably estimate $\bcmb$ from these analyses if $\alpha$ is not well determined (and indeed $\alpha$ is not exactly known for current experiments).

\citet{Minami:2019:rot} pointed out a solution to this problem: the authors noted that one could break the degeneracy between $\bcmb$ and $\alpha$ arising from a measurement of $\bhi$ by using the fact that nearby polarized galactic foreground emission is generally not rotated by cosmic birefringence but is only affected by $\alpha$ (hereafter, we refer this method to as the foreground-based $\alpha$-calibration method). \citet{Minami:2020:planck} recently applied the foreground-based $\alpha$-calibration method to the latest Planck polarization data \citep{P18:main} and obtain a constraint on the isotropic cosmic birefringence of $\bcmb=0.35\pm 0.14$\,deg (68\% C.L.). \bds{However, some knowledge of the foreground $EB$ correlation is needed for this measurement; \cite{clark} recently pointed out that the indeterminate sign of the foreground $EB$ spectrum could complicate this analysis.}

Since we, for many models, expect $\bdelt$ to be non-zero, we can construct a new estimator for birefringence that is immune to instrument miscalibration. Consider making a measurement, using the large scale ($\ell \lessapprox 20$) $EB$ power spectrum, of the angle $\blo$ by which specifically the polarization signal from reionization has been rotated. If we now calculate the difference between the measured polarization rotation angle of recombination polarization $\bhi$ from smaller scales and the reionization polarization angle $\blo$, we can see that the instrument miscalibration angle $\alpha$ cancels: $\bhi-\blo = (\alpha+\bcmb)-(\alpha+\bri) = \bcmb-\bri = \bdelt$. Put differently, a measurement of both the reionization and recombination polarization rotation angles $\bhi$ and $\blo$ allows us to break the degeneracy with instrument angles and cleanly determine the cosmological birefringence difference $\bdelt$. 
Using WMAP data, \citet{Komatsu:2008:WMAP5} were previously able to constrain the birefringence angles on both large and small scales, corresponding to roughly $\blo$ and $\bhi$, respectively. However, no previous studies have addressed the possibility of using the difference between $\blo$ and $\bhi$ (or an equivalent joint analysis) as a probe of birefringence and ALPs that is immune to polarization angle systematic errors. 

We also note that, when compared with the foreground-based $\alpha$-calibration method, our method is sensitive to Galactic foregrounds in a different way. The foreground-based $\alpha$-calibration method treats Galactic foregrounds as a source of information rather than just a contaminant, and therefore depends to a larger extent on the modeling of the correlation structure of Galactic foregrounds, in particular on an assessment of the intrinsic $EB$ spectrum of the Galactic foreground components. Our method, on the other hand, does not directly rely on Galactic foregrounds as a source of information about birefringence, and instead regards them as only a contaminant to be removed (with multifrequency component separation methods). Our method therefore requires a different type of knowledge, and \bds{perhaps less knowledge}, of the foregrounds' $EB$ spectra and other spatial correlations.

\emph{2. Birefringence tomography using reionization polarization:} if we can obtain an independent determination of $\bcmb$, for example from the foreground-based $\alpha$-calibration method or from a very well-calibrated instrument, combining this with a determination of $\bdelt$ can give further insight into the redshift-dependence of any new physics causing the birefringence. \footnote{Even a measurement of $\bdelt$ alone could give insights into the redshift-dependence of physics causing any birefringence.} We can again consider ALPs as an example. If $\bcmb$ is found to be non-zero (using either a well-calibrated instrument or the foreground-based $\alpha$-calibration method), from this measurement alone it is not clear at which redshifts significant changes in the ALP field took place. However, adding a measurement of $\bri$ (or $\bdelt$) can give us more information. If we find that $\bri \approx \bcmb$ (or equivalently $\bdelt \approx 0$), this indicates that the ALPs only began to oscillate and dilute significantly after reionization. If, on the other hand, we find that $\bri \ll \bcmb$ (i.e., that $\bdelt \approx \bcmb$), we conclude that the ALPs must have begun to oscillate before reionization. Since the ALP field begins to oscillate when $H$ becomes $\sim m$, the former case corresponds to a lower-mass ALP with $m \ll H_{\rm reio} $, the latter to a higher mass ALP with $m \gg H_{\rm reio}$ (here $H_{\rm reio}$ is the Hubble parameter at reionization). \bds{This example illustrates how including a measurement of the reionization birefringence can constrain the redshift dependence of any birefringence effect and hence provide insights into its physical origin.}

\section{Detailed Methodology and Forecasting}
In the following section we outline our method in more detail.

Denoting the CMB Stokes $Q/U$ parameters generated at reionization and recombination as $Q^{\rm reio}$/$U^{\rm reio}$ and $Q^{\rm recomb}$/$U^{\rm recomb}$, respectively, the observed $Q/U$ parameters measured after rotation (denoted by primed variables) are given by:
\al{
    Q'\pm\iu U' = \sum_{x=\rm reio,recomb}(Q^x\pm \iu U^x)\E^{\pm2\iu \theta_x } 
    \,. 
}
The $E$- and $B$-modes are obtained by a spin-$2$ spherical harmonic transform of the above Stokes Q/U parameters  \citep{Kamionkowski:1996:eb,Zaldarriaga:1996:EBdef}.  From the above equation, the observed $E$- and $B$-modes are given by (see e.g.  \citealt{Zhao:2014:biref}): 
\al{
    E'_{lm} &= \sum_{x=\rm reio,recomb}[E^x_{lm}\cos2 \theta_x -B^x_{lm}\sin2 \theta_x]
    \,, \label{Eq:E:rot} \\ 
    B'_{lm} &= \sum_{x=\rm reio,recomb}[B^x_{lm}\cos2 \theta_x + E^x_{lm}\sin2 \theta_x]
    \,, \label{Eq:B:rot}
}
where $E^{\mathrm{reio}}/B^{\mathrm{reio}}$ and $E^{\mathrm{recomb}}/B^{\mathrm{recomb}}$ again indicate the relevant polarization generated at reionization and recombination respectively. The $EB$ power spectrum is then
\al{
    C^{EB\prime}_l = \sum_{x=\rm reio,recomb}\frac{C^{EE,x}_l-C^{BB,x}_l}{2}\sin4 \theta_x
    \,. 
}
Since the rotation angles, $\alpha$ and $\beta_x$, are typically $\mC{O}(0.1)\,$degrees or less, we assume $\alpha,\beta_x\ll 1$ and hence $\theta_x\ll 1$ such that 
\al{
    C^{EB\prime}_l \simeq \sum_{x=\rm reio,recomb}2 \theta_x (C^{EE,x}_l-C^{BB,x}_l)
    \,. \label{Eq:EB}
}
In addition to the above signal, the observed $EB$ power spectrum may also contain contaminants such as Galactic foregrounds. 
\tn{For a single parameter, $\theta_{\rm recomb}$, we can constrain $\theta_{\rm recomb}$ from observed $EB$, $EE$ and $BB$ spectra as was done in previous analyses. if we consider two parameters, $\theta_{\rm reio}$ and $\theta_{\rm recomb}$, we note that they can be constrained separately without significant degeneracy, because the reionization signal dominates at very large angular scales ($l\alt 10$) while the recombination signal is a dominant source of polarization anisotropies on smaller scales. Intuitively, if we constrain the birefringence angle using very large scale polarization modes, we obtain primarily $\blo = \alpha+\bri$; using smaller-scale polarization data, we constrain mainly $\bhi = \alpha+\bcmb$. As stated previously, differencing the two angles we can cancel out $\alpha$ and obtain $\bdelt$.} 

\tn{To see the feasibility of our new approach,} we will evaluate the expected constraints on $\bdelt$ with the Fisher information matrix formalism. Assuming that, at the field level, the observed CMB $E$- and $B$-modes have a multivariate zero-mean Gaussian distribution, the Fisher information matrix is given by \citep{Tegmark:1996:Fisher}:
\al{
	\bR{F}_{ij} = \sum_{l} \fsky\frac{2l+1}{2}
		\Tr\left(\bR{C}^{-1}_l\PD{\bR{C}_l}{p_i} \bR{C}^{-1}_l\PD{\bR{C}_l}{p_j}\right)\bigg|_{p_i=p_{i,\rm fid}}
	\,, 
}
where $\fsky$ is the fractional area of the observed sky, $p_i$ are the parameters to be constrained, $p_{i,\rm fid}$ are the fiducial parameter values used in our forecast, and the covariance of the $E$- and $B$-modes is given by:
\al{
	\bR{C}_l \equiv \Mat{ \hC_l^{EE} & \hC_l^{EB} \\ \hC_l^{EB} & \hC_l^{BB} }
	\,. 
}
The above covariance contains the observed angular power spectra of $EE$, $EB$ and $BB$ which contain the signal, noise and, where indicated, residual Galactic foregrounds after component separation (note that spectra with hats, unlike spectra with primes, also include noise). 
We consider two parameters, $\blo$ and $\bhi$. The derivative of the covariance with respect to $\theta_x$ only contains the off-diagonal elements. The above Fisher matrix can be simplified as follows:
\al{
	\bR{F}_{\theta_x\theta_y} 
	&= \sum_{l} \fsky(2l+1)4(C^{EE,x}_l-C^{BB,x}_l)(C^{ EE,y}_l-C^{ BB,y}_l)
	\notag \\
	&\quad \times \frac{\hC^{EE}_l\hC^{BB}+(\hC^{EB}_l)^2}{[\hC_l^{EE}\hC_l^{BB}-(\hC^{EB}_l)^2]^2}
	\,. 
}
We choose the fiducial values of $\blo$ and $\bhi$ to be zero and also assume the fiducial value of the $EB$ spectrum to be $\hC_l^{EB}=0$, yielding:
\al{
	\bR{F}_{\theta_x\theta_y} 
	&= \sum_{l} \fsky(2l+1)\frac{4(C^{EE,x}_l-C^{BB,x}_l)(C^{EE,y}_l-C^{BB,y}_l)}{\hC_l^{EE}\hC_l^{BB}}
	\,. \label{Eq:fisher}
}
The above Fisher matrix is equivalent to that derived by assuming a Gaussian distribution for $\hC_l^{EB}$ instead of for the CMB multipoles, although this coincidence is not general.

We note that the constraint on $\bhi$ is much tighter than that on $\blo$, since for all realistic experiments the number of modes at $l \geq 20$ is much larger than at $l \leq 20$. Due to this (and the previously mentioned lack of degeneracy), we are justified in approximating $\bhi$ as exactly determined so that the constraint on $\bdelt = \bhi - \blo$ is, to a very good approximation (which we have tested\footnote{\bds{We can also verify this directly by computing the Fisher matrix of \eq{Eq:fisher} with two angle parameters, $\blo$ and $\bhi$, and comparing $\sigma(\blo)$ obtained from this Fisher matrix with that obtained without marginalizing $\bhi$. We find that the increase in $\sigma(\blo)$ is negligibly small as expected.}}), simply set by the error on the reionization signal angle $\blo$. \bds{This approximation has the advantage that the} error on $\blo$ is also the relevant quantity to forecast for birefringence tomography if $\bhi$ has been determined by another method.

We therefore only consider $\blo$ as a free parameter for our forecasts, approximating $\bhi$ as fixed by the argument above.
We compute the $1\sigma$ constraint on $\bdelt$ as $\sigma(\bdelt)\approx \sigma(\blo)\approx 1/\sqrt{F_{\blo \blo}}$ where the Fisher matrix is obtained from \eq{Eq:fisher}. In evaluating this Fisher matrix, we obtain the reionization contribution to the $E$-mode power spectrum by differencing the $E$-mode power with $\tau = 0.06$ and $\tau =0$ using the following combination: $C^{EE,\rm reio}_l = C^{ EE}_l(\tau = 0.06) - e^{-2 \times 0.06}C^{EE}_l(\tau = 0)$; we additionally null the very small contribution above $l >20$. We further assume that any $B$-modes from inflationary gravitational waves \bds{or patchy reionization} are negligible in our forecasts and approximate $C^{BB,\rm reio}_l=0.$
The fiducial values of the LCDM cosmological parameters we use are $h=0.675, \Omega_b h^2=0.022, \Omega_c h^2=0.122, \tau=0.06, A_s=2 \times 10^{-9}, n_s=0.965$. In the following analysis, we assume $f_{\rm sky}=0.7$ throughout, although to obtain constraints over a different sky area the forecast errors can by simply scaled with a factor$\sqrt{0.7/f_{\rm sky}}$. \tn{In our baseline calculation, we assume $l_{\rm min}=2$.}

\section{Forecast Results}

\begin{figure}
	\includegraphics[width=1.\columnwidth]{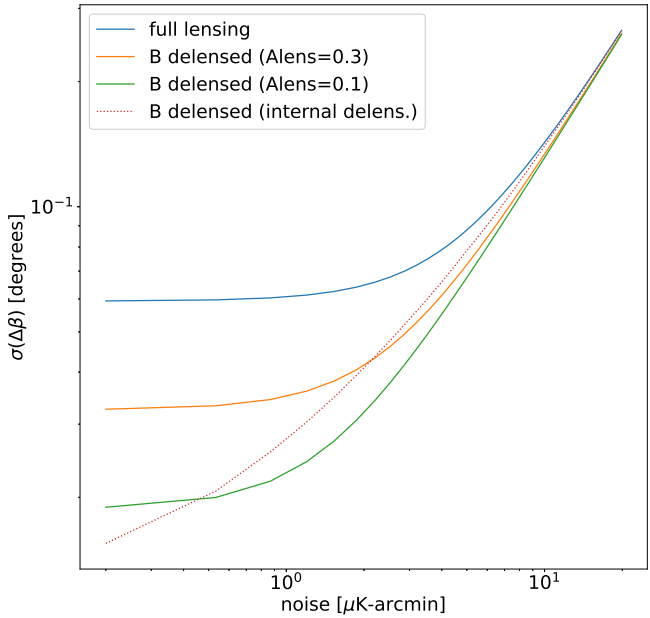}
    \caption{Constraints on the birefringence angle difference as a function of CMB polarization noise level, with different delensing efficiencies assumed (we also assume $f_\mathrm{sky}=0.7$.) Solid lines of different colours assume different constant delensing efficiencies; the dotted line labelled `internal delens.' assumes delensing using lensing measurements by the same experiment. These results show that for future CMB satellites reaching noise levels of a few $\mu$K-arcmin, competitive constraints on the birefringence angle difference of order $0.05$ degrees can be achieved. Note that here any Galactic foreground residuals are not included in the forecasts.}
    \label{fig:basicFigure}
\end{figure}

\begin{figure}
	\includegraphics[width=1.\columnwidth]{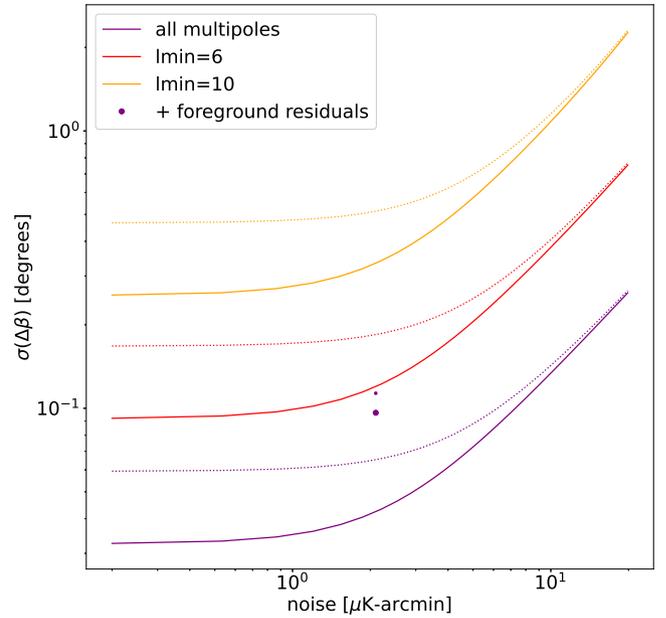}
    \caption{Variation of the constraints in Figure 1 with the minimum multipole used in the analysis and with the presence of foreground residuals. Solid lines assume lensing B-power has been reduced by a factor 0.3, dotted lines assume no delensing (and both neglect foreground residuals). Purple colour indicates a standard analysis using the full multipole range. It can be seen that low multipoles are important for deriving tight constraints, although even if we must exclude $\ell<6$, competitive constraints of $\sim 0.1$ degrees can be achieved. To show the effects of foregrounds, the two dots indicate the impact of foreground residuals on the errors for a LiteBIRD-like experiment with a $2\mu$K-arcmin noise level; values are calculated by including in the errors the post-component-separation foreground residual power for LiteBIRD obtained in \citet{Errard_2016} (the lower dot assumes delensing such that lensing B-power has been reduced by a factor 0.3). We note that this is a conservative estimate, since we expect the $EB$ power spectrum to be less contaminated by foregrounds than $BB$, so that less aggressive foreground cleaning methods could be employed.}
    \label{fig:lminFigure}
\end{figure}

We show the expected constraints on the recombination -- reionization cosmic birefringence angle difference $\bdelt$ in Fig.~\ref{fig:basicFigure} as a function of instrumental noise in polarization. 
\tn{In this section, we assume that the observed CMB power spectra contain only signal and instrumental noise. The impact of foregrounds is discussed in the next section.}

The results also depend on the the amount of delensing, or lensing B-mode removal, that has been applied to the $B$-mode polarization map in order to reduce the scatter. Different surveys are pursuing different delensing strategies and so we include several options in our forecast plots. Delensing methods in which a tracer from a different survey is used to estimate the lensing $B$-modes, or external delensing, result in an approximately fixed amount of residual lensing $B$-mode power (which enters into the error calculation for our birefringence constraint). Curves are therefore shown for three different fixed levels of external delensing: no delensing (blue), delensing resulting in a residual lensing $B$-mode power that is 30\% of its original value (orange, labelled $\mathrm{Alens}=0.3$), and delensing resulting in a 10\% residual power (green, labelled $\mathrm{Alens}=0.1$). Another possibility is to delens using a lensing map reconstructed from the same CMB experiment (internal delensing); the dotted line shows the results assuming internal delensing for an arcminute-beam experiment as a function of noise level.

It can be seen that for upcoming or planned CMB satellites, such as LiteBIRD, PICO or CORE, all of which can achieve few-micro-Kelvin level noise, competitive constraints on birefringence can be achieved. For example, considering a polarization noise level of $2 \mu$K-arcmin (which is similar to the noise of the LiteBIRD CMB channels) we find that constraints on $\bdelt$ of order $\sigma(\bdelt) \sim 0.05$ degrees are achievable. In particular, with typical external delensing, we obtain a constraint of 0.02-0.03 degrees; without delensing, we forecast constraints of 0.06 degrees. 

These results compare favorably with other methods for constraining birefringence, although, as discussed in the next section, we caution that a more detailed foreground treatment in our forecasts is needed for an exact comparison. In particular, using the foreground-based $\alpha$-calibration method for LiteBIRD, \citet{Minami:2020:litebird} forecast $\sigma(\beta)\simeq 0.1\,$deg with CMB-dominated bands and $\sigma(\beta)\simeq 0.06\,$deg with all of the possible bands of LiteBIRD.
\bds{Our constraints should, in addition, be significantly better than the constraints on $\beta$ without self-calibration from future experiments. For instance, \citet{Sekimoto:2021:litebird-design} reports a design goal for LiteBIRD of $\sim0.05\,$deg but this requirement is expected to be very hard to achieve without $EB$ self-calibration; since our forecast constraints are significantly tighter than the angle errors on $\alpha$ derived from a hardware calibrator or an astrophysical source, our method for measuring $\bdelt$ appears well motivated. (Of course, even if instrument angles were perfectly calibrated, comparing birefringence of reionization and recombination signals would still give interesting constraints on the redshift origin of any observed birefringence.)}

In Figure \ref{fig:lminFigure} we examine the dependence of these constraints on the minimum usable multipole $l$, assuming either no delensing or delensing by a factor of $0.3$. It can be seen that, as reionization signals dominate on large scales, much of the constraining power arises from low multipoles, as expected. However, even if we must exclude $l<6$, interesting constraints of order $\sigma(\bdelt) \sim 0.1$ degrees can be obtained.

\section{Foreground considerations}

\begin{figure}
	\includegraphics[width=1.12\columnwidth]{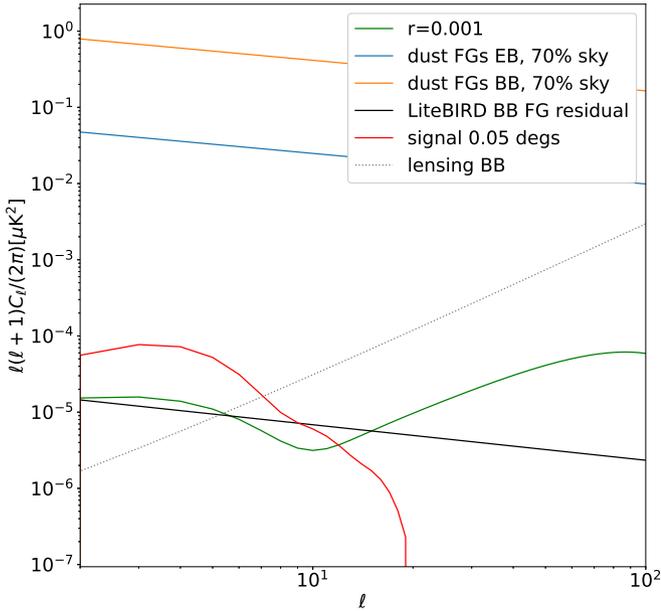}
    \caption{Foreground estimates and comparison to the expected signals for both birefringence and inflationary gravitational wave polarization. Red solid line: expected reionization $EB$ power spectrum signal arising from birefringence of $0.05$ degrees. Green solid line: inflationary $B$-mode polarization power for $r=0.001$. Orange solid line: 130 GHz foreground polarization $BB$ power inferred from Planck measurements. Blue solid line: conservative estimate of the 130 GHz $EB$ dust polarization assuming the $EB$ power is $3\%$ of the dust $EE$ power (saturating the current Planck upper limit). It can be seen that the (red) $EB$-power signal is typically larger than the (green) $BB$-power signal for $r=0.001$ at low multipoles, while the (blue) $EB$ dust power is significantly lower than the (orange) foreground power for a $BB$ measurement.}
    \label{fig:example_figure}
\end{figure}


The measurement of $\blo$ appears challenging since it arises from large angular scale ($\ell \leq 20$) polarization, which suffers from significant foreground contamination from both galactic dust and synchrotron sources.

Nevertheless, we note that there are several factors which make the measurement of $\blo$ and hence $\bdelt$ less challenging than a measurement of inflationary gravitational wave $B$-modes on similarly large scales.

We note first that some of the primary methods for performing multifrequency cleaning of low-$l$ polarization will be pixel-based methods such as \citet{Errard_2016, Stompor_2016, Errard_2019}; in these methods, cleaning $E$-modes should work comparably well to cleaning $B$-modes since the frequency scaling of the CMB is perfectly known (neglecting systematics such as bandpass errors) and the $E$-mode cosmic variance does not enter. We can therefore estimate the scale of the foreground challenge by comparing the levels of expected foregrounds and signals in the $EB$ power spectrum to those in the $BB$ power spectrum. In our initial argument we assume that we can neglect any frequency-dependence in the instrument angle error, which may not be true; we briefly revisit this complication at the end of this section. We will also focus our initial discussion on dust foregrounds.

First, we note that the large-scale reionization $EB$-power signal \bds{arising from $\bri = 0.05 \sim \sigma(\bri)$} is significantly larger than the reionization $BB$-power corresponding to $r=0.001$, a typical target of next-generation CMB experiments such as LiteBIRD (which aims to detect inflationary $B$-modes at this level from the reionization feature as well as the recombination feature). This is shown quantitatively in Fig.~\ref{fig:example_figure}: at low multipoles, the red $EB$-signal lies well above the green line, which corresponds to $B$-mode power at the level of $r=0.001$.

Second, measurements indicate that dust foreground levels in the $EB$-spectrum are at least an order of magnitude lower than foreground power in the $BB$-spectrum.
To illustrate this in Fig.~\ref{fig:example_figure}, we first plot the 130 GHz $BB$ foreground levels on $70$\% of the sky from \citet{delabrouille}, obtained from Planck polarization measurements. Deriving an estimate for the dust-$EB$ levels is more difficult, since dust-foreground-$EB$ power has not yet been detected and currently only upper limits exist. However, we may make an approximate and conservative estimate of the $EB$ power from dust by assuming that the $EB$ dust power spectrum is given by $\sim 3\%$ of the dust $EE$ power spectrum (which we, in turn, assume to equal $2 \times$ the $BB$ foreground power spectrum); this level of $EB$ foregrounds approximately equals the $1-$sigma upper limit obtained from Planck \citep{planckdust}. The result is shown in blue in Fig.~\ref{fig:example_figure}; as previously indicated, even this conservative estimate, which saturates current upper limits, is much smaller than the foreground $BB$ power shown in orange. We note that an alternative upper limit was obtained by \citet{clark}, which is consistent with the Planck-derived result we quote here. \footnote{It should be noted that, since in our method (unlike in the foreground-based $\alpha$-calibration method) we do not rely on foregrounds for our measurement, we are free to choose an analysis mask where $EB$-foregrounds are minimized; since the $EB$-dust spectrum could be both positive or negative, potentially a mask could be chosen that sets the $EB$-foreground bias close to zero.}

With a larger signal and, even estimating conservatively, a significantly smaller foreground level than $BB$ measurements targeted by the same experiments, determining $\bri$ from future satellites' \bds{$EB$ spectra} appears possible despite foreground complexity.

Nevertheless, extensive further analysis will be required to test this simple argument more quantitatively. In particular, an additional complication we have not considered in our previous discussion is a scenario in which the instrumental angle error varies rapidly with frequency. This could potentially produce complications when applying foreground cleaning algorithms, since the angle-error-induced leakage of $E$- to $B$-mode signal can no longer be described with one single angle. However, we note that in multifrequency component separation algorithms, the linear combination weights at frequencies far from the CMB dominated frequencies from $90-150$ GHz are generally small since the foreground emission is very large at these frequencies. Therefore the impact of an already-small CMB angle error at very high or low frequencies may well be negligible. In addition, if one is able to ensure that the same frequency-cleaning weights are applied to both the high-$l$ recombination polarization analysis and the low-$l$ reionization analysis, one can preserve the cancellation of the effective angle error when differencing reionization and recombination measurements. We defer a detailed investigation of this and other complications and mitigation strategies (see, e.g., \citealt{Verges}) to future work.


A related question which we briefly discuss is the following: will the required foreground cleaning significantly weaken birefringence constraints by increasing the effective noise level? To investigate this, we assume the foreground cleaning prescription of \citet{Errard_2016}. In this method the foregrounds (both galactic dust and synchrotron) are cleaned with a map-level parametric component separation approach. The result is an increase in the effective noise due to the presence of residual foreground power; the foreground residual power for a LiteBIRD-like experiment is shown in Fig.~\ref{fig:example_figure} with a black solid line. We use these post-foreground cleaning residuals to revisit our forecasts for a LiteBIRD-like noise level of $2\mu$K-arcmin, adding the foreground residual power to the power spectra used to derive covariances. We find that this degrades the constraints on $\bdelt$ to some extent, as shown by the purple dots in Fig.~\ref{fig:lminFigure}, but that the resulting bounds remain competitive, giving constraints on $\bdelt$ of order $0.09$ degrees. We note that this is a conservative estimate, because -- as discussed previously -- we expect the $EB$ power spectrum to be less contaminated by foregrounds than $BB$, so that less aggressive foreground cleaning methods with a lower noise penalty could be employed.

While, as stated previously, much more work on foreground removal is required, we therefore expect our forecast constraints to be approximately correct despite the complexities of foreground cleaning.

\section{Conclusions}
In this paper we propose a new method of searching for cosmic birefringence by differencing the angle constraints obtained from CMB recombination and reionization signals; a non-zero difference angle $\bdelt$ can arise quite generically in models that produce non-zero cosmic birefringence. We point out that this difference angle measurement has two key advantages: first, its measurement is insensitive to instrumental angle errors, which cancel in the difference; second, by combining with other techniques to determine the CMB birefringence, it provides tomographic information that can provide insight into the redshift dependence of any new physics that is responsible for rotating polarization.

Performing forecasts for future experiments, we find that for experiments such as LiteBIRD, competitive constraints of order $\sigma(\bdelt) \sim 0.05$ degrees appear achievable, with the exact performance depending on the details of delensing and foreground cleaning assumed.

Future work in this area will require a detailed analysis of foreground mitigation for the birefringence measurement from large-scale reionization polarization signals. In this work, we have sketched out a simple argument for why these foreground challenges should be tractable (effectively, the signal is significantly larger and the relevant foregrounds are significantly smaller than for the standard $B$-mode power spectrum analyses targeted by experiments such as LiteBIRD). Nevertheless, we do not discuss complications such as a strong frequency dependence of the instrumental angle error or possible systematic errors in the $EB$ spectrum measurement; we defer a detailed analysis of such challenges to future work.


\section*{Acknowledgements}

We are grateful to Eiichiro Komatsu, Colin Hill, Anthony Challinor, Josquin Errard and Francesca Chadha-Day for comments on a draft of this manuscript and for helpful discussions; we also thank Max Abitbol for useful discussions. Some of the results in this paper have been derived using public software, CAMB \citep{Lewis:1999bs}. 
BDS acknowledges support from the European Research Council (ERC) under the European Unions Horizon 2020 research and innovation programme (Grant agreement No. 851274) and an STFC Ernest Rutherford Fellowship. 
TN acknowledges support from the JSPS KAKENHI Grant Number JP20H05859 and World Premier International Research Center Initiative (WPI), MEXT, Japan.
For numerical calculations, this paper used resources of the National Energy Research Scientific Computing Center (NERSC), a U.S. Department of Energy Office of Science User Facility operated under Contract No. DE-AC02-05CH11231.

\section*{Data Availability}
The data that support the findings of this study are available from the corresponding author, BDS, upon reasonable request.



\bibliographystyle{mnras}
\bibliography{cite} 




\appendix




\bsp	
\label{lastpage}
\end{document}